\pdfoutput=1
\documentclass[namedreferences]{SolarPhysics}
\usepackage[optionalrh]{spr-sola-addons} 
\usepackage{graphicx}        
\usepackage{color}           
\usepackage{url}             

\newcommand{\doiurl}[1] {\url {http://dx.doi.org/#1}}
\newcommand{\adsurl}[1] {\url {http://adsabs.harvard.edu/abs/#1}}

\newcommand{\etal}{{\it et al.}}

\newcommand{\apjl}{    {\it Astrophys. J. Lett.}}

\begin{document}

\begin{article}

\begin{opening}

\title{A Change of Solar He\,{\sc ii} EUV Global Network Structure of the Transition Region as an Indicator of Geo-Effectiveness of Solar Minima}

\author{L.~\surname{Didkovsky}$^{1}$\sep
        J.B.~\surname{Gurman}$^{2}$
       }
\runningauthor{L.~Didkovsky, and J.B.~Gurman} \runningtitle{A Change of EUV Global Network Structure}

   \institute{
                $^{1}$ Space Sciences Center, University of Southern California, Los Angeles, USA \\
                    email: \url {leonid@usc.edu}\\
                $^{2}$ NASA Goddard Space Flight Center, Greenbelt, MD 20771\\
                     email: \url {Joseph.B.Gurman@nasa.gov}\\
             }

\begin{abstract}
Solar activity during 2007\,--\,2009 was very low, causing anomalously low thermospheric density. A comparison of solar extreme ultraviolet (EUV) irradiance in the He\,{\sc ii} spectral band (26 to 34 nm) from the \textit{Solar Extreme ultraviolet Monitor} (SEM), one of instruments on the \textit{Charge Element and Isotope Analysis System} (CELIAS) onboard of the \textit{Solar and Heliospheric Observatory} (SOHO) for the two latest solar minima showed a decrease of the absolute irradiance of about 15 $\pm$ 6\,\%  during the solar minimum between Cycles 23 and 24 compared with the Cycles 22/23 minimum when a yearly running mean filter was used. We found that some local, shorter-term minima including those with the same absolute EUV flux in the SEM spectral band show a larger concentration of  spatial power in the global network structure from the 30.4~nm SOHO \textit{Extreme ultraviolet Imaging Telescope} (EIT) images for the local minimum of 1996 compared with the minima of 2008\,--\,2011. We interpret this larger concentration of  spatial power in the transition region's global network structure as a larger number of  larger area features on the solar disk. Such changes in the global network structure during solar minima may characterize, in part, the geo-effectiveness of the solar He\,{\sc ii} EUV irradiance in addition to the estimations based on its absolute levels.

\end{abstract}
\keywords{Solar Extreme ultraviolet irradiance; Solar cycle; Solar minimum}
\end{opening}

\section{Introduction}
     \label{S-Introduction}

The absolute solar EUV irradiance in the He\,{\sc ii} spectral band of 26 to 34~nm from the SEM was 15 $\pm$ 6$\,\% $ lower during the Cycles 23/24 solar minimum compared to the Cycles 22/23 minimum \cite{Didkovsky2010}. This measured lower EUV irradiance combined with the modeled effects, \textit{e.g.} due to the amount of CO$_{2}$ in the Earth's atmosphere \cite{Roble89,Qian06} caused anomalously low thermospheric density during the most recent solar minimum \cite{Solomon10}. A number of other investigations showed a global change in the upper atmosphere during the Cycles 23/24 solar minimum (\opencite{Lastovicka06}, \citeyear{Lastovicka08}) including the thermosphere \cite{Emmert04,Emmert10}, \textit{e.g.} using  satellite-drag data \cite{Emmert08}. The decrease in the solar irradiance  was detected in different spectral bands; see, \textit{e.g.}, the long-term trend in the total solar irradiance \cite{Frohlich09} or the low solar EUV irradiance in 2008 \cite{Chamberlin09}. \inlinecite {Solomon11} found that the thermospheric density showed the lowest values  at any  time in the past 47 years. Both the NCAR Thermosphere-Ionosphere-Electrodynamics General Circulation Model \cite{Roble88} and SEM EUV data \cite{Didkovsky2010} showed good agreement between  thermospheric density changes from 1996 to 2008 and the changes in the solar 26 to 34 nm EUV irradiance. \inlinecite{Woods10} found that the 2008 to 1996 modeled decrease in the XUV (0 to 15~nm) irradiance ratio was even larger, up to 35 \%. It remains unclear, however, whether some other change on the Sun is responsible for such a significant decrease in the solar irradiance, \textit{e.g.} in the EUV. For example, the oppositely directed effects of the decreased number of polar coronal holes in 2007 \cite{Kirk09} and the presence of some large-area, low-latitude coronal holes in 2008 discussed by \inlinecite{Woods10} make it impossible to conclude whether coronal holes can explain the long-term decrease of  solar EUV irradiance during the 2007\,--\,2009 minimum. Another question is about the geo-effectiveness of the absolute solar EUV irradiance during the periods of the same levels of the daily averaged SEM irradiance, \textit{e.g.} in 1996 and 2010: is the geo-effectiveness of the EUV irradiance the same? For example, \inlinecite{Solomon11} mentioned that  geomagnetic activity became extremely low in mid-2008 and remained low until late 2009, spanning about 1.5 years of some of the quietest solar-terrestrial conditions ever observed. The authors found that 2009, with an annual average Ap $= 4$, was slightly lower in annual average density than 2008 although the solar-irradiance conditions were apparently similar.

\inlinecite{McIntosh11a} used a ``watershed segmentation'' applied to a 30.4~nm image from the EIT \cite{Delaboudiniere1995} to determine the distribution of cell radii
in the central portion of the image, up to $0.6$ of the solar radius. The segment radii have quite a wide range, about 11 to 40 Mm with the largest frequency of appearance about ten for 24 Mm cells from the 10 April 2008 EIT image. \inlinecite{McIntosh11b} used this method to calculate and compare the mean Transition Region (TR) network scale for the available EIT 30.4 nm images from 1996 to 2011; see, \textit{e.g.}, the lower panel on Figure 1 of \inlinecite{McIntosh11b}. The authors found a significant decrease in the network scale during the Cycles 23/24 solar minimum compared with the Cycles 22/23 minimum.

In this work we analyze spatial power spectra from SOHO/EIT 30.4 nm images for different periods of low solar activity including those for the Cycles 22/23 and Cycles 23/24 solar minima. Rather than using histograms of the cell radii \cite{McIntosh11a} or the mean TR network scale \cite{McIntosh11b}, we compare the spatial power of the images in the whole ranges of the image radii and spatial frequencies. In addition to the method of Magnetic Range of Influence (MRoI) \cite{McIntosh06, Leamon09} the proposed approach may be used to explore how the change in the spatial scale of the TR network affects the Earth's thermosphere.

\section{Data Observations}
\label{S-Data Observations}

Our analysis of  periods of low solar activity is based on EUV images obtained with the EIT in the He\,{\sc ii} band. Data selected for this analysis (Table 1) represent the deepest local solar minima detected in the He\,{\sc ii} band (last column) by SOHO/CELIAS/SEM during last two solar cycles, 22/23 and 23/24, three local minima in 2010 detected by SEM and \textit{Extreme Ultraviolet Spectrophotometer} (ESP: \opencite{Didkovsky2012}) onboard the \textit{Solar Dynamics Observatory Extreme ultraviolet Variability Experiment} (SDO/EVE: \opencite{Woods2012}), and one local minimum in 2011. In addition to the EIT images, we also examined  spatial power spectra in the 30.4 nm band from the \textit{Atmospheric Imaging Assembly} (AIA: \opencite{Lemen12}) onboard  SDO for comparison of the minima of 2010 and 2011.

\begin{table}
\caption{EIT images selected for  periods of low solar activity including some local minima for Solar Cycles 22/23 (April and May 1996) and 23/24 (November 2008). Image properties are compared using the sum of on-disk signal in the units of Data Numbers [DN] and the mean disk signal (third and fourth columns). Daily averaged absolute EUV SEM flux in the 26\,--\,34~nm bandpass is shown in the last column for a reference.}
\begin{tabular}{ccccc} 
\hline
  Date & File Name  & Sum of C$_{\rm eff}$, & Mean Signal, & SEM Flux,  \\
       &   efz*     & 10$^{8}$ [DN] & [DN] & 10$^{10}$ [ph\,cm$^{-2}$\,sec$^{-1}$]\\
\hline
27\,Apr\,96 & 19960427.033225 & 3.99 & 380.8 & 1.08 \\
27\,Apr\,96 & 19960427.092959 & 3.91 & 373.3 & 1.08 \\
28\,Apr\,96 & 19960428.003126 & 3.93 & 376.5 & 1.08 \\
28\,Apr\,96 & 19960428.062000 & 3.93 & 374.5 & 1.09 \\
28\,Apr\,96 & 19960428.120829 & 3.91 & 372.9 & 1.09 \\
28\,Apr\,96 & 19960428.175716 & 3.91 & 373.2 & 1.09 \\
04\,Jun\,96 & 19960604.192903 & 1.62 & 309.5 & 1.17 \\
04\,Jun\,96 & 19960604.201208 & 1.60 & 305.8 & 1.17 \\
05\,Jun\,96 & 19960605.170704 & 1.58 & 301.9 & 1.18 \\
08\,Jun\,96 & 19960608.025934 & 1.53 & 291.0 & 1.19 \\
27\,Nov\,08 & 20081127.011936 & 0.65 & 123.9 & 0.95 \\
27\,Nov\,08 & 20081127.071936 & 0.65 & 124.2 & 0.95 \\
27\,Nov\,08 & 20081127.191936 & 0.65 & 124.4 & 0.95 \\
28\,Nov\,08 & 20081128.011935 & 0.65 & 123.5 & 0.94 \\
28\,Nov\,08 & 20081128.121937 & 0.65 & 123.1 & 0.94 \\
28\,Nov\,08 & 20081128.191937 & 0.65 & 123.4 & 0.94 \\
29\,Nov\,08 & 20081129.011934 & 0.64 & 121.1 & 0.93 \\
29\,Nov\,08 & 20081129.071936 & 0.64 & 121.5 & 0.93 \\
29\,Nov\,08 & 20081129.121938 & 0.63 & 121.0 & 0.93 \\
29\,Nov\,08 & 20081129.201937 & 0.64 & 122.2 & 0.93 \\
14\,May\,10 & 20100514.011937 & 0.57 & 107.8 & 1.01 \\
14\,May\,10 & 20100514.071938 & 0.58 & 111.3 & 1.01 \\
14\,May\,10 & 20100514.131938 & 0.57 & 108.0 & 1.01 \\
14\,May\,10 & 20100514.191938 & 0.56 & 107.6 & 1.01 \\
15\,May\,10 & 20100515.071939 & 0.56 & 107.5 & 1.01 \\
15\,May\,10 & 20100515.191938 & 0.56 & 106.4 & 1.01 \\
16\,May\,10 & 20100516.011939 & 0.56 & 106.1 & 1.00 \\
16\,May\,10 & 20100516.071938 & 0.56 & 105.9 & 1.00 \\
16\,May\,10 & 20100516.131938 & 0.56 & 106.5 & 1.00 \\
16\,May\,10 & 20100516.191938 & 0.56 & 106.0 & 1.00 \\
26\,Aug\,10 & 20100826.011940 & 0.57 & 108.2 & 1.09 \\
26\,Aug\,10 & 20100826.131938 & 0.56 & 106.8 & 1.09 \\
27\,Aug\,10 & 20100827.011941 & 0.56 & 107.0 & 1.09 \\
28\,Aug\,10 & 20100828.131939 & 0.55 & 105.3 & 1.08 \\
19\,Dec\,10 & 20101219.011938 & 0.53 & 101.8  & 1.16 \\
20\,Dec\,10 & 20101220.012011 & 0.54 & 102.2 & 1.14 \\
20\,Dec\,10 & 20101220.131940 & 0.53 & 101.4  & 1.14 \\
21\,Dec\,10 & 20101221.011939 & 0.53 & 101.2  & 1.13 \\
21\,Dec\,10 & 20101221.131938 & 0.54 & 102.2 & 1.13 \\
22\,Dec\,10 & 20101222.011941 & 0.54 & 102.2 & 1.14 \\
23\,Dec\,10 & 20101223.011938 & 0.53 & 101.1  & 1.15 \\
23\,Dec\,10 & 20101223.131937 & 0.53 & 101.1  & 1.15 \\
24\,Dec\,10 & 20101224.011938 & 0.53 & 100.7  & 1.14 \\
24\,Dec\,10 & 20101224.131938 & 0.52 & 99.1  & 1.14 \\
25\,Dec\,10 & 20101225.011941 & 0.51 & 98.1  & 1.13 \\
25\,Dec\,10 & 20101225.131939 & 0.51 & 97.8  & 1.13 \\
19\,May\,11 & 20110519.131941 & 0.43 & 82.6 & 1.28 \\
20\,May\,11 & 20110520.011940 & 0.43 & 82.7 & 1.26 \\
21\,May\,11 & 20110521.131941 & 0.43 & 82.1 & 1.26 \\
24\,May\,11 & 20110524.131620 & 0.44 & 83.3 & 1.27 \\
\hline
\end{tabular}
 \vspace{-0.03\textwidth}
\end{table}
Table 1 shows a steady decrease of the sum of the on-disk pixel signals (third column) and the mean level of these signals (fourth column) compared with the changes of the daily averaged absolute 26\,--\,34 nm SEM flux (last column). This decrease may indicate a degradation of the EIT thin-film aluminum filter. We discuss this and the other forms of degradation in Sections 3\,--\,5. The numbers in the third and fourth columns are corrected for changes in EIT CCD pixel quantum efficiency (QE) based on the visible light calibration lamp (CL) tests.

\section{Data Reduction} 
      \label{S-Data Reduction}

Raw EUV 30.4~nm counts from the original EIT files (Table 1) were first converted to effective data numbers [DN] by subtracting corresponding dark images.
\begin{equation}    
C_{\rm eff}(i,j)=C(i,j) - C_{\rm dark}(i,j),
\end{equation}
where $i$ and $j$ are the indices of the image array. The center of the solar disk and its radius from each image file header were used to determine the image pixels outside the disk area and set these pixel signals to zero.
As an example, Figure 1 shows EIT He\,{\sc ii} solar disk image for 27 April 1996 in the units DN. Note, all pixel signals outside the disk are set to zero.
  \begin{figure}[ht]
   \begin{center}
   \begin{tabular}{c}
\includegraphics[height=9.0 cm]{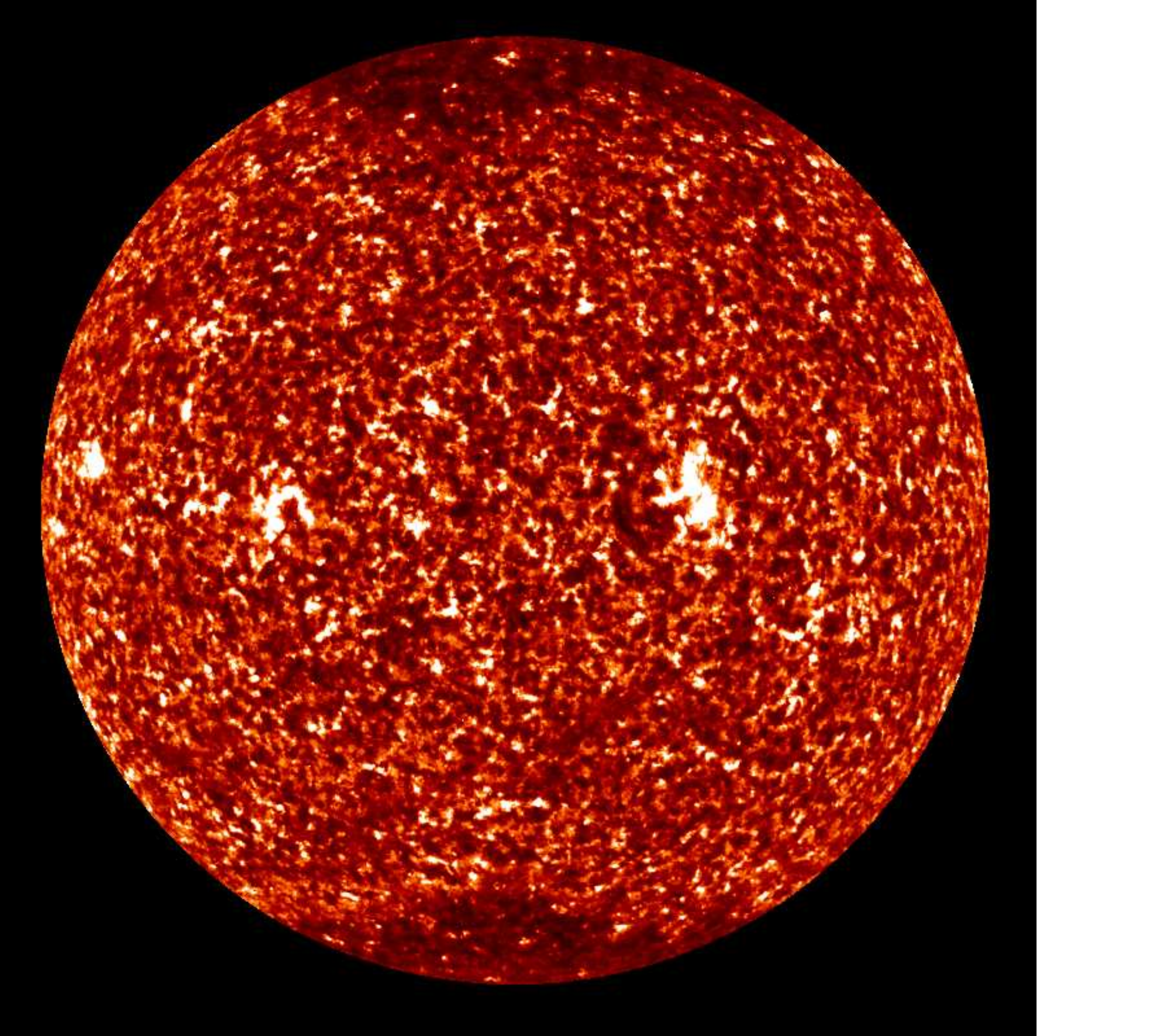}
   \end{tabular}
   \end{center}
   \caption[Figure 1]
   { \label{fig:Figure 1}
An example of the image reduced according to Equation (1) for 27 April 1996 shows effective counts $C_{\rm eff}$ inside the solar disk using a color scale and zero counts outside the disk. }
   \end{figure}
Then non-zero pixels were organized as an one-dimensional vector
 \begin{equation}       
C_{\rm eff}(k)=C_{\rm eff}(i,j)
\end{equation}
for the $k$ pixels within the disk area. Bright one-pixel spikes related to the CCD impacts from high-energy particles were replaced with the mean DN for the images. Our goal was to calculate the ratios of the signal distribution as a function of the spatial frequency in the power spectra of such one-dimension sequences for different groups of files (dates) in Table 1 taken for the low levels of solar activity.

\subsection{Pixel Degradation and Power Spectra} 
      \label{S-Pixel Degradation}

EIT data files available from \url{umbra.nascom.nasa.gov/eit/eit-catalog} are not corrected for the degradation.
A comparison of the sum of the pixel effective counts (third column in Table 1) for about the same levels of daily averaged SEM absolute flux (last column in Table 1) in April 1996 and August 2010 shows a decrease of roughly a factor of seven. This change of the total solar-disk image brightness is associated with the degradation of the thin-film aluminum filter and/or EIT reflective optics. Similar degradation of the thin-film aluminum filter was determined for the ESP channels based on the daily and weakly calibrations. This degradation of the whole brightness, \textit{e.g.} the mean level, may be easily corrected by a single degradation coefficient [$D(\lambda,t)$] which balances the fluxes for different time intervals. The mean level $\rm mean(C_{\rm eff}(k))$ is subtracted
 \begin{equation}       
C_{k}=C_{\rm eff}(k)- \rm mean(C_{\rm eff}(k))
\end{equation}
in the data prepared for calculation of the power spectra [$\Psi_{f}$].
 \begin{equation}       
\Psi_{f}=\sum_{k=0}^{K-1} C_{k}\mathrm{e}^{\mathrm{-i 2} \pi f k/K}
\end{equation}
where $f=0,..,K-1$ is the spatial frequency array, $K$ is the number of pixels within the solar disk image stored as one-dimensional vector (Equation (2)), and $C_{k}$ is either absolute signal variations (Equation (3)) or relative variations
 \begin{equation}       
C_{k}=\frac{C_{\rm eff}(k)- {\rm mean(C_{\rm eff}(k))}}{\rm mean(C_{\rm eff}(k))}
\end{equation}

Usually, the degradation related coefficient [$D(\lambda,t)$] is a function of wavelength [$\lambda$] and the time [$t$] of exposure of the filter/optics to solar radiation.
For our analysis of the solar EIT images in the same 30.4~nm spectral band, the wavelength does not affect the ratios analyzed and the degradation coefficient [$D(t)$] is a function of time only. This coefficient is the same to correct the degradation of effective counts or the power spectrum based on these effective counts
   \begin{equation}     
C_{1}(k)/(C_{2}(k) D(t))=\Psi_{1}/(\Psi_{2} D(t))
\end{equation}
because $D(t)$ is not a function of the pixel number $k$.
Thus, the power spectrum of the degraded data with decreased levels of the spectral density would be shifted up to match the level of the less degraded curve for the highest spatial frequency. This shift of the whole spectrum brings the ratios between these curves to the unity for the highest spatial frequency, which provides
a convenient approach to compare ratios for different spatial frequencies and different time intervals.  The effect of the degradation is shown in Figure 2 (a) where the whole bottom curve for August 2010 represents some spatial power lower than the upper curve for April 1996. Figure 2 (b) shows the same curves after the bottom curve was shifted up to match the top curve in Figure 2 (a) for the highest spatial frequencies. The multiplicative coefficient $D(1996/2010)$ was determined based on the ratio of the mean power for $10^{3}$ points of the power spectra in the range of frequencies
 \begin{equation}       
\Delta f=f(K/2) -f(K/2-1000)
\end{equation}

  \begin{figure}[ht]
  \centerline{\hspace*{0.015\textwidth}
  \includegraphics[width=0.515\textwidth,clip=]{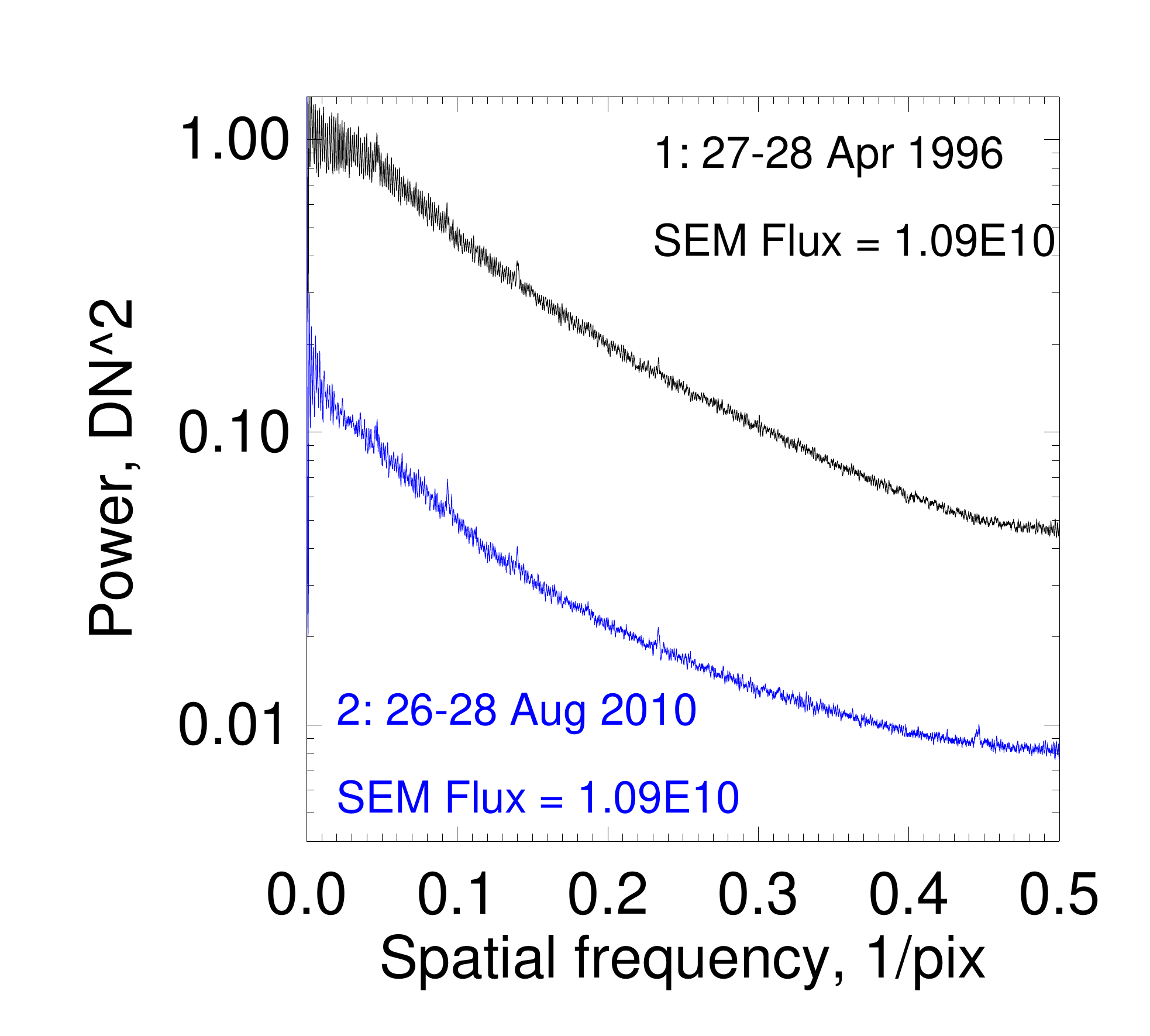}
               \hspace*{-0.03\textwidth}
               \includegraphics[width=0.515\textwidth,clip=]{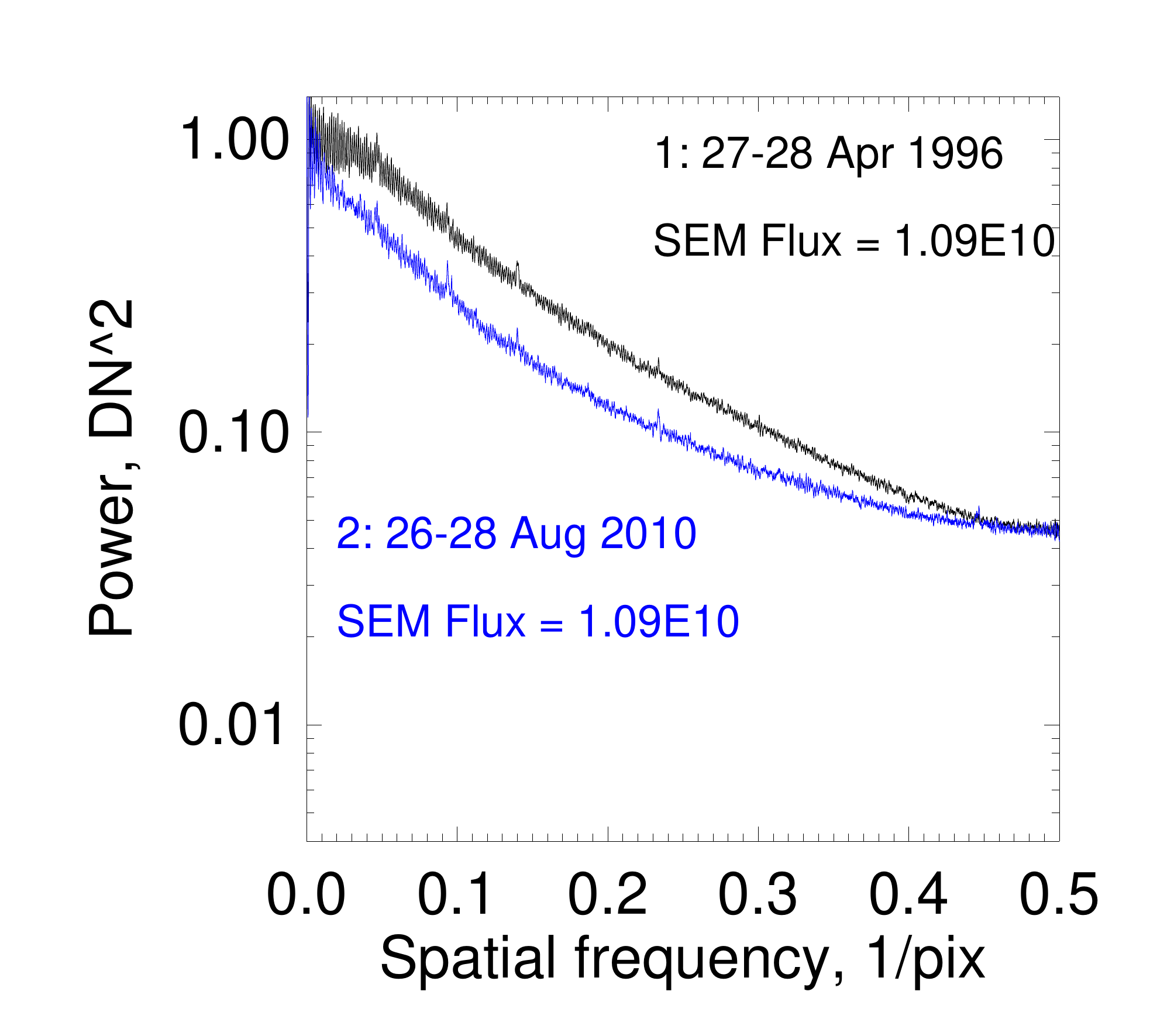}
              }
     \vspace{-0.42\textwidth}   
     \centerline{\Large \bf     
      \hspace{0.0 \textwidth}  \color{black}{(a)}
      \hspace{0.415\textwidth}  \color{black}{(b)}
         \hfill}
   \vspace{0.42\textwidth}
   \caption[Figure 2]
   { \label{fig:Figure 2}
Running-mean curves through the power-spectra data for six files in April 1996 marked as $1$ and for four files in August 2010 marked as $2$ (see also Table 1). Note small differences between the curves for each time interval. (a) The curves for 2010 show a lower spectral density due to the throughput degradation $D(1996/2010)$ of the 2010 images compared to the reference images for 1996. For the examples shown in Figures 2 (a) and (b) the spectra were calculated based on the absolute amplitudes (Equation 3). (b) Same as in (a), but with the levels of the two spectra matching each other for the initial highest spatial frequency by multiplying the spectra for 2010 by $D(1996/2010)= 5.6$. Note, the coefficient $D(1996/2010)$, used to match the curves for the initial frequency, represents a small portion ($10^3$ points) of the power-spectra data and thus is different from a degradation ratio based on the sum of the pixel signals or the mean number (third and fourth columns in Table 1).
}
   \end{figure}
Another degradation factor that we analyze is related to a possible change of the QE of the EIT CCD pixels as a function of time between detector bakeouts. EIT test data  \url {umbra.nascom.nasa.gov/eit/eit_guide/euv_degradation} show some changes of pixel QE ratios. These changes were determined using flat-field measurements provided by a visible-light CL. CL data used in our analysis are shown in Table 2.
\begin{table}
\caption{EIT calibration-lamp images used to determine the CCD pixel changes of the QE. The images were selected to provide the closest time intervals to the data shown in Table 1. }
\begin{tabular}{cc} 
\hline
  Date & File Name  \\
       &   efz*     \\
23\,May\,1996 & 19960523.153137  \\
24\,Jun\,1996 & 19960624.204011  \\
26\,Jul\,2008 & 20080726.040008  \\
29\,Nov\,2008 & 20081129.040610  \\
15\,May\,2010 & 20100515.040053  \\
28\,Aug\,2010 & 20100828.040849  \\
18\,Dec\,2010 & 20101218.040902  \\
20\,May\,2011 & 20110520.040645  \\

\end{tabular}
 \vspace{-0.03\textwidth}
\end{table}
CCD QE changes based on the CL files (Table 2) were calculated as pixel ratios for each pair of CL images which corresponds to the timing of the group of files used for the comparison (Table 1). The CL image data were first converted to the effective signals (CL minus dark) and then transferred to one-dimensional arrays for the pixels from the inside of the disk areas, as shown in Equation (2). These ratios [$CL_{1}(k)/CL_{2}(k)$] were then used to correct the pixel signals
 \begin{equation}       
C_{\rm corr}(k)=C_{2}(k) \frac{CL_{1}(k)}{CL_{2}(k)}
\end{equation}
where indices $1$ and $2$ correspond to the earlier and later file dates. As an example, Figure 3 shows the ratios for two CL dates, 23 May 1996 and 28 Aug 2010.
  \begin{figure}[h]
   \begin{center}
   \begin{tabular}{c}
   \includegraphics[height=7.0 cm]{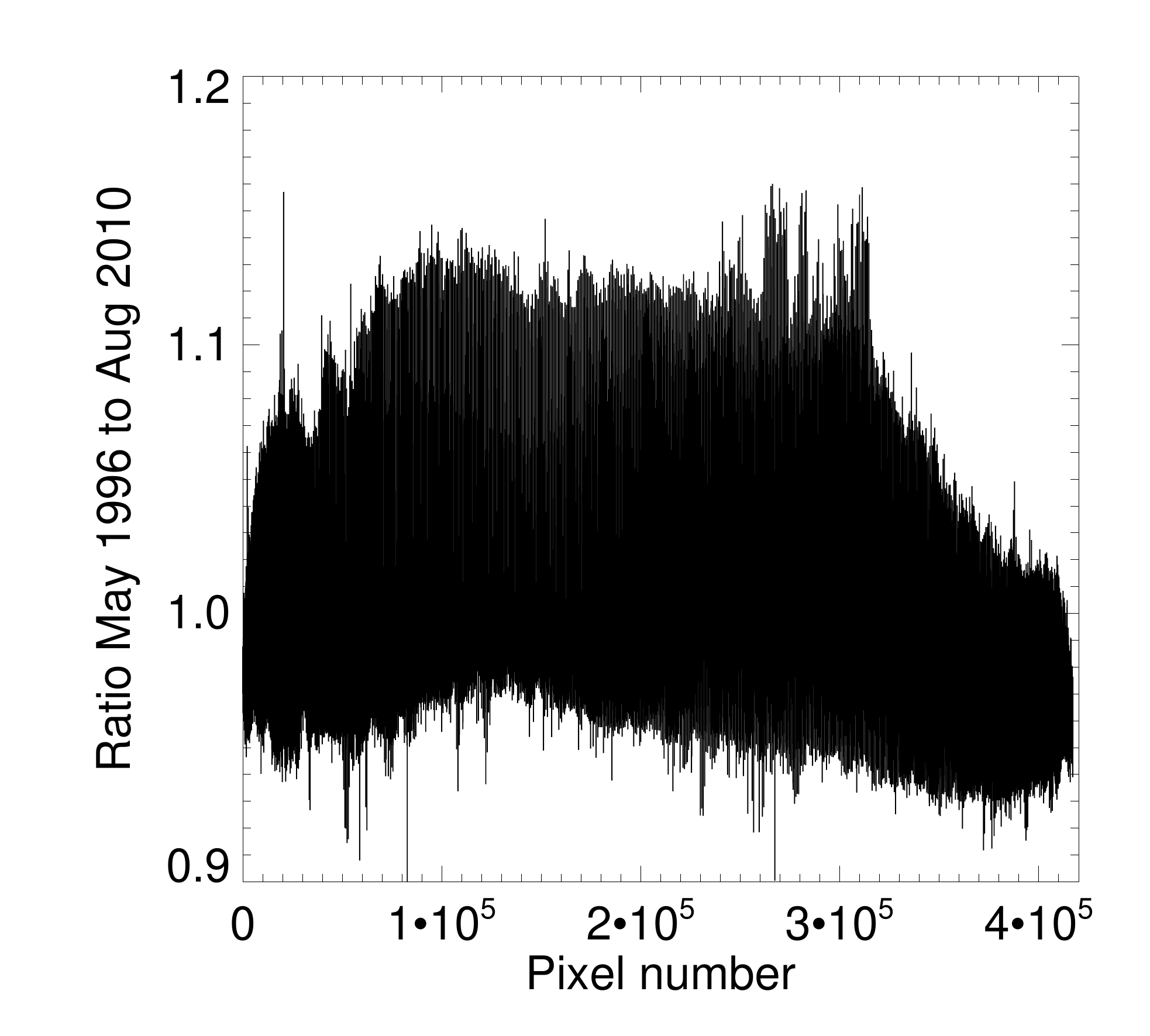}
   \end{tabular}
   \end{center}
   \caption[Figure 3]
   { \label{fig:Figure 3}
 CL pixel ratios for the pixels within the solar-disk area organized as a one-dimensional vector for the images taken on 23 May 1996 and 28 August 2010 (Table 2).  }
   \end{figure}

In addition to the QE changes registered with the visible-light CL, we analyze whether changes in the power-spectra ratios (next section) might be due to the CCD QE changes in the 30.4 nm EIT spectral band. A wavelength-dependent degradation of the telescope optics, thin-film filters and CCD could produce degradation similar to that detected by EVE and AIA during the SDO mission. For example, degradation of the thin-film filters decreases the output signals due to the deposition of some hydrocarbons on the filter area. The amount of this deposition is a function of the filter exposure time to the solar radiation and is wavelength dependent.

\section{Comparison of Power Spectra for Different Time Intervals with Quiet-Sun Conditions} 
      \label{S-Comparison of Power Spectra}

Power spectra (Equation (4)) were calculated using one-dimension pixel array, which contains relative signal variations (Equation (5)) after their correction for the visible light CL ratios. For each power spectrum for the solar images (Table 1) a running mean filter curve over the spectrum amplitude [$\Psi_{f}$] (Equation (4)) was calculated. This low-pass filter is based on the running mean procedure (IDL procedure $\sf {median}$) with the same integration window of 355 data (spatial frequency) points. The total number of frequency points [$f$] in the spatial frequency range from zero to 0.5 pix$^{-1}$ for the EIT spectra is about 2$^{19}$. The number of pixels [$k$] within the solar-disk area is slightly smaller and the remaining data array points are filled with its mean number. The 355 data points from the power spectrum used for the integration in the running-mean procedure correspond to a range of spatial frequencies of about 6.8 $^{-4}$ pix$^{-1}$ for the EIT spectra. These running-mean curves for a number of images within
each group of files (Table 1) were averaged and the ratios for each pair of the averaged curves together with the standard deviation (STD) were calculated.

\subsection{EIT Power Spectra Ratios}
    \label{S-EIT Power Spectra Ratios}

 We have compared power spectra for five different levels of solar activity: four with relatively short time intervals separating them and one with a long time interval. The ratios for short time intervals (Figure 4) are for April 1996 \textit{vs.} June 1996 (diamonds), November 2008 \textit{vs.} May 2010 (filled circles), May 2010 \textit{vs.} August 2010 (squares), and December 2010 \textit{vs.} May 2011 (triangles).
  \begin{figure}[ht]
   \begin{center}
   \begin{tabular}{c}
   \includegraphics[height=7.0 cm]{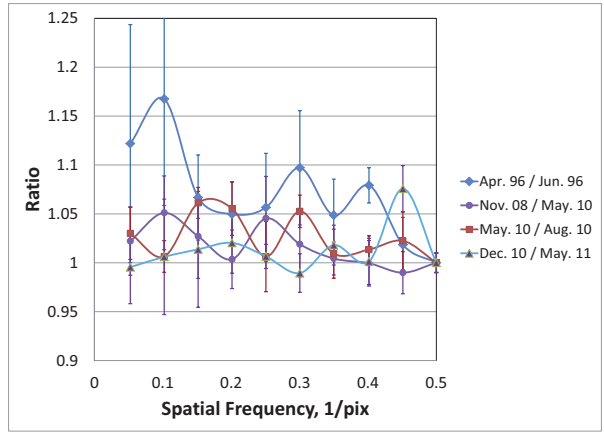}
   \end{tabular}
   \end{center}
   \caption[Figure 4]
   { \label{fig:Figure 4}
 Power spectra ratios for April 1996 \textit{vs.} June 1996 (diamonds), November 2008 \textit{vs.} May 2010 (filled circles), May 2010 \textit{vs.} August 2010 (squares), and December 2010 \textit{vs.} May 2011 (triangles). }
   \end{figure}
Figure 4 shows that the ratios in the spatial frequency range of 0.05 to 0.3 pix$^{-1}$ were largest for the first pair of time intervals (April 1996 \textit{vs.} June 1996) except for the frequency of 0.2 pix$^{-1}$, and smaller for the three later pairs of intervals. This may be interpreted as faster changes of the He\,{\sc ii} spatial structure of the EIT images in 1996 and about the same small changes in 2008\,--\,2011.

For the long time interval, we have chosen two groups of files with about the same level of EUV (30.4 nm) SEM flux of 1.085 $10^{10}$ ph\,cm$^{-2}$\,sec$^{-1}$ for April 1996 and August 2010. The ratios for these time intervals are shown in Figure 5.
  \begin{figure}[h]
   \begin{center}
   \begin{tabular}{c}
   \includegraphics[height=7.0 cm]{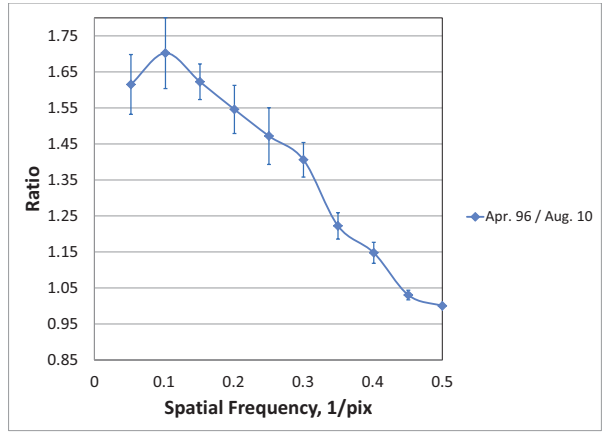}
   \end{tabular}
   \end{center}
   \caption[Figure 5]
   { \label{fig:Figure 5}
 Power spectra ratios for April 1996 \textit{vs.} August 2010 (diamonds) with about the same level of SEM EUV flux. }
   \end{figure}
Figure 5 shows that the ratios between April 1996 and August 2010 power spectra at low spatial frequencies are significant, several times larger than the STD errors.

\subsection{Comparison of SOHO/EIT and SDO/AIA Power Spectra Ratios}
    \label{S-EIT and AIA comparison}

Since the beginning of the SDO mission in February 2010, there have been a few time intervals with quiet-Sun conditions. The first two were in May and August 2010. These two time intervals are interesting to compare due to very different operation time and ratios of possible degradation for SOHO/EIT (15 years of observations) and SDO/AIA (less than a year in 2010). Data reduction for the AIA images was similar to that described for EIT images with the difference of substantially larger size of one-dimensional array:
2$^{23}$ compared to 2$^{19}$ for EIT and no additional correction for degradation. AIA images were taken from the AIA website:
\url{www.lmsal.com/get_aia_data/} and SolarSoft IDL programs were used to download the data.
For each of the days analyzed, we downloaded six files with 60-minute cadence. The results of this comparison are shown in Table 3.

\section{Discussion} 
      \label{S-Discussion}

Power spectra for two time intervals in April 1996 and August 2010 (Figure 2 (b)) show significant changes of the ratios for the spatial frequencies between 0.05 and 0.4 pix$^{-1}$ (Figure 5). One may think that these ratio differences reflect the degradation of the EIT 30.4~nm channel not corrected by the proposed data reduction. We have performed, tested, and compared five ``instrumental'' approaches to prove that the ratios between the spatial structure of EUV images in 1996 and 2010 are not related to throughput degradation.

The first is a correction of EIT degraded flux by shifting the data from more recent time intervals to match the spatial frequencies at and below 0.5 pix$^{-1}$ (Figure 2 (a) and (b)). This operation corrects the image degradation related to the decreased reflectivity of the optics and/or decreased transmission of the filters.

The second approach was to compare the ratios for four short time intervals (Figure 4). Figure 4 shows that the ratios for April 1996 \textit{vs.} June 1996 are larger than for the three other pairs of the time intervals. The averaged ratios for these four pairs are shown in the second column of Table 3. If the ratios were related to the degradation, one would expect a steady decrease of the ratios toward the last (Dec. 10 / May. 11) pair. In contrast to this, the ratios for the second pair (filled circles) is smaller than for the two last pairs (squares and triangles).
Formal STD errors for the averaged ratios are shown in the third column. They show that for all time intervals, except the second, the averaged ratios are larger than unity.
\begin{table}
\caption{Averaged ratios and STDs for four pairs of compared time intervals.}
\begin{tabular}{ccccc} 
\hline
  Compared         & EIT Ratio  & EIT STD & AIA ratio & AIA STD \\
  Time Intervals   &            &         &           &         \\
\hline
 Apr 1996 \textit{vs.} Jun 1996   & 1.070   & 0.05      &            &         \\
 Nov 2008 \textit{vs.} May 2010   & 1.016   & 0.03      &            &         \\
 May 2010 \textit{vs.} Aug 2010   & 1.035   & 0.02      &   1.035    &  0.01   \\
 Dec 2010 \textit{vs.} May 2011   & 1.031   & 0.02      &   1.031    &  0.02   \\
\hline
\end{tabular}
 \vspace{-0.03\textwidth}
\end{table}

The third test was to compare the ratio results from two EUV 30.4 nm data channels, SOHO/EIT with its 15 years of operation, and SDO/AIA (Table 3) with less than one year of operation in 2010. A comparison of averaged ratios for EIT and AIA (third and fourth rows in Table 3) shows the match of the ratios. This match of the EIT and AIA averaged ratios with significantly different time of operation and rate of degradation may be used to prove that the ratios for the third and fourth compared groups reflect the change in the image structures. Note, Table 3 compares the averaged ratios in the whole range of the spatial frequencies not affected by the differences in the EIT and AIA spatial resolution, 2.6 and 0.6 [arcsec pix$^{-1}$] or 1.9 and 0.4 [Mm pix$^{-1}$].

The fourth test was based on known effects of degradation for SDO/EVE and SDO/AIA instruments. The degradation reduces the detector/pixel contrast as a function of both exposure time of the instrument to the solar beam and wavelengths. As a result of different EIT and AIA mission (exposure) time, one might expect different ratios for EIT and AIA power spectra (third and fourth rows in Table 3), which is not the case. In addition to the averaged ratio comparison in Table 3, we have tested the ratios for the pair shown in Figure 5 but with a number of different contrast coefficients. The pixel-contrast variations for the most recent images (August 2010) used to compare with the April 1996 images (Figure 5) were modified [$C_{\rm mod}(k)$] using Equations (9) and (10). Equation (9) was used for the pixel signal larger than the mean [$C_{\rm mean}(k)$] for the one-dimensional array. Equation (10) was used for the pixel signals smaller than the mean [$C_{\rm mean}(k)$], where

 \begin{equation}
C_{\rm mod}(k)= C(k) \cdot m
\end{equation}
or
\begin{equation}
C_{\rm mod}(k)= C(k) / m
\end{equation}
where $m$ is the contrast increase/decrease coefficient. Figure 6 shows a set of ratio curves for the original ratios (bottom curve) and for the power spectra using modeled images with modified contrast.
  \begin{figure}[ht]
   \begin{center}
   \begin{tabular}{c}
   \includegraphics[height=7.0 cm]{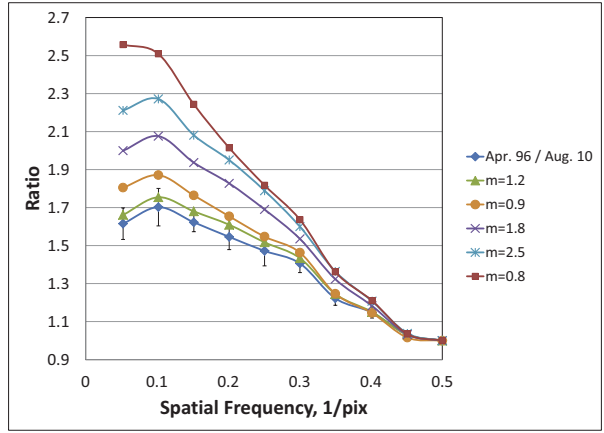}
   \end{tabular}
   \end{center}
   \caption[Figure 6]
   { \label{fig:Figure 6}
 A comparison of power spectra ratios for the original pair (diamonds) of time intervals April 1996 \textit{vs.} August 2010 and the ratios for five modified contrast curves above it for $m$ equal to 1.2 (triangles), 0.9 (filled circles), 1.8 (crosses), 2.5 (stars), and 0.8 (squares). Note, modification of the contrast was applied to the most recent (more degraded) series of the EIT images.}
   \end{figure}
Figure 6 shows that the changes of the pixel contrast of the most recent images in a wide range of $m$, from 0.8 to 2.5, do not decrease the ratios compared to the original curve. This test proves that the change of the image pixel-to-pixel contrast associated with the degradation of EIT August 2010 images compared to the less degraded April 1996 images does not lead to the smaller spatial differences of the ratios, which could be close to unity for the similar spatial structures.

Finally, a fifth test consisted of adding two realizations of random noise, with a smaller and an order of magnitude larger amplitudes to the pixel signals of either August 2010 or April 1996 images. The results of added pixel-to-pixel random noise shows some increase of the spatial power at high spatial frequencies (0.35 to 0.5) pix$^{-1}$ but does not affect the power (and ratios) at lower spatial frequencies.

These five ``instrumental'' tests lead us to conclude that the measured ratios based on the power spectra differences for 1996 \textit{vs.} 2010 with about the same EUV SEM flux in the 30.4~nm band (Figure 5) are not related to a component of EIT throughput degradation that has not been accounted for by our data reduction algorithm.

If the 1996 \textit{vs.} 2010 differences (Figure 5) in the spatial ratios are not due to ``instrumental'' sources, what solar sources could be responsible?
To determine which solar feature(s) is (are) responsible for the changes of the ratios three tests were performed.
First, we compared the ratios from the power spectra with polar regions removed from the solar-disk images to exclude polar coronal holes, visible, \textit{e.g.}, in Figure 1. The result of these modified spectra was practically the same as for the full-Sun images.

Second, we compared the ratios for only the central portions of the solar disk with a limiting  radius equal to $0.75$ of the original radius. In addition to removing the polar regions from the disk images, this procedure also removes  areas close to the limb with image radii between $0.75$ and $1.00$. The result was the same, with no more than $3\,\%$ change to the ratios.

Finally, we stretched the solar image for 28 August 2010 at 13:19:39 (Figure 7 (a)) to change the spatial dimensions of the solar features on the disk (Figure 7 (b)).
  \begin{figure}[ht]
  \centerline{\hspace*{0.015\textwidth}
  \includegraphics[width=0.495\textwidth,clip=]{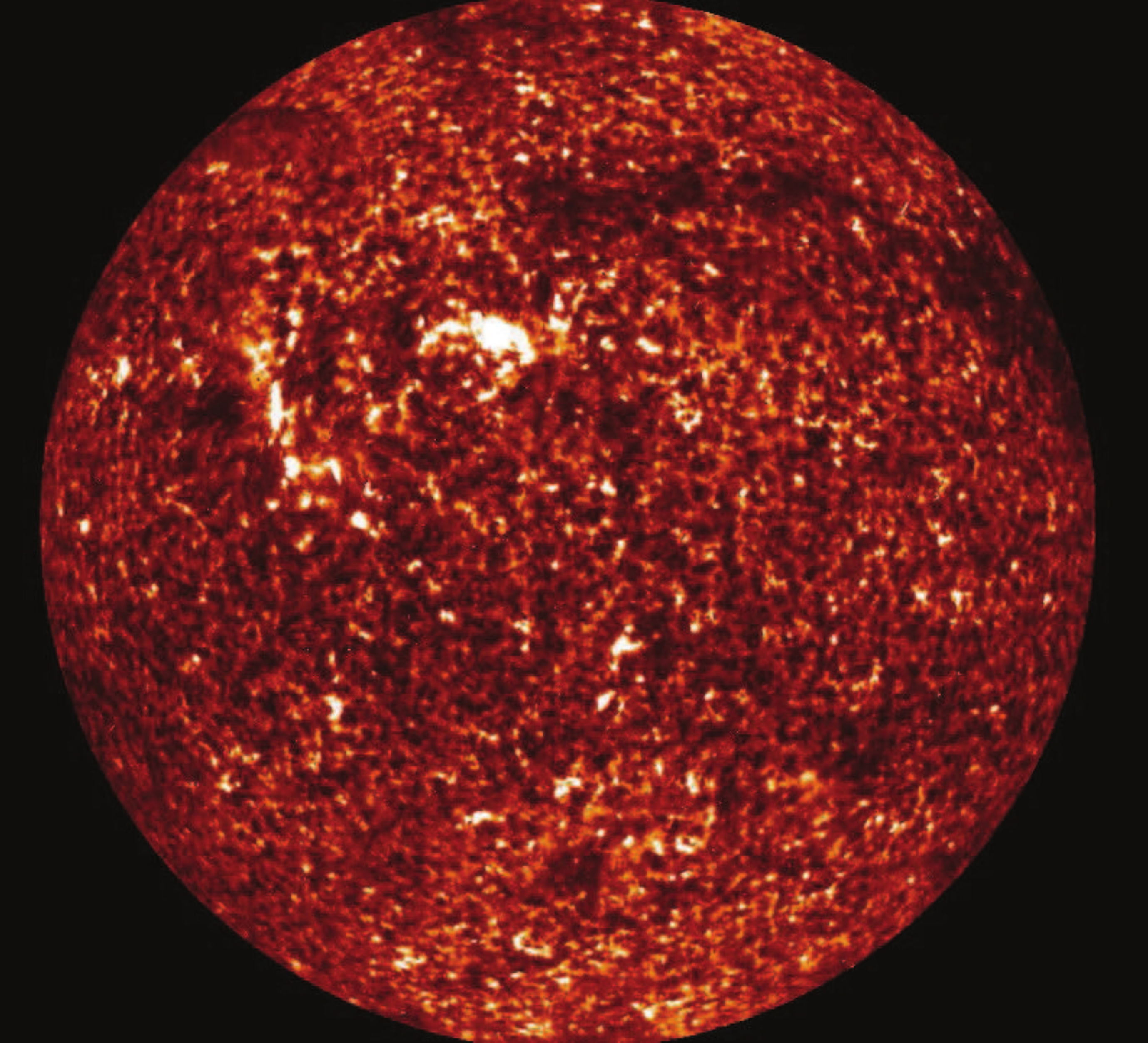}
               \hspace*{-0.03\textwidth}
               \includegraphics[width=0.495\textwidth,clip=]{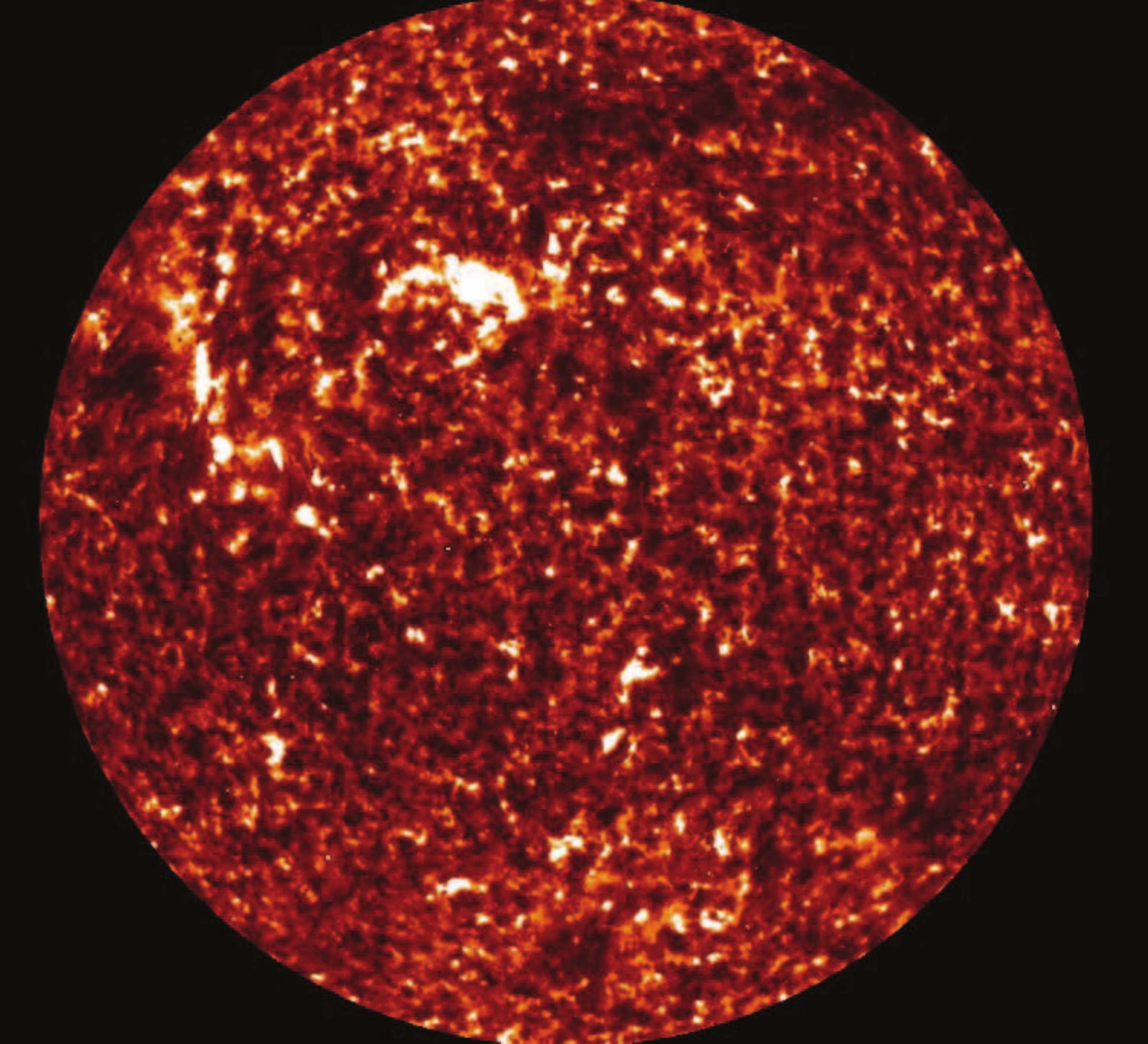}
              }
     \vspace{-0.42\textwidth}   
     \centerline{\Large \bf     
      \hspace{0.0 \textwidth}  \color{white}{(a)}
      \hspace{0.415\textwidth}  \color{white}{(b)}
         \hfill}
   \vspace{0.42\textwidth}
   \caption[Figure 7]
   { \label{fig:Figure 7}
(a) Original EIT 30.4 nm solar image for 28 August 2010 at 13:19:39. (b) The same image but first stretched from $1024 \times 1024$ pixels to $1280 \times 1280$ pixels and then saved with the same format as the left image.
}
   \end{figure}
The stretching procedure is to increase the image size by 1.25 times and then use the same image area as in the original image. The increase of the image dimensions increases the spatial details of the He\,{\sc ii} structure on the image (compare Figure 7 (a) and (b)). Saving the same area on the stretched image causes the removal of a small outer ring compared to the original image. The result of this procedure is shown in Figure 8,
  \begin{figure}[h]
   \begin{center}
   \begin{tabular}{c}
   \includegraphics[height=7.0 cm]{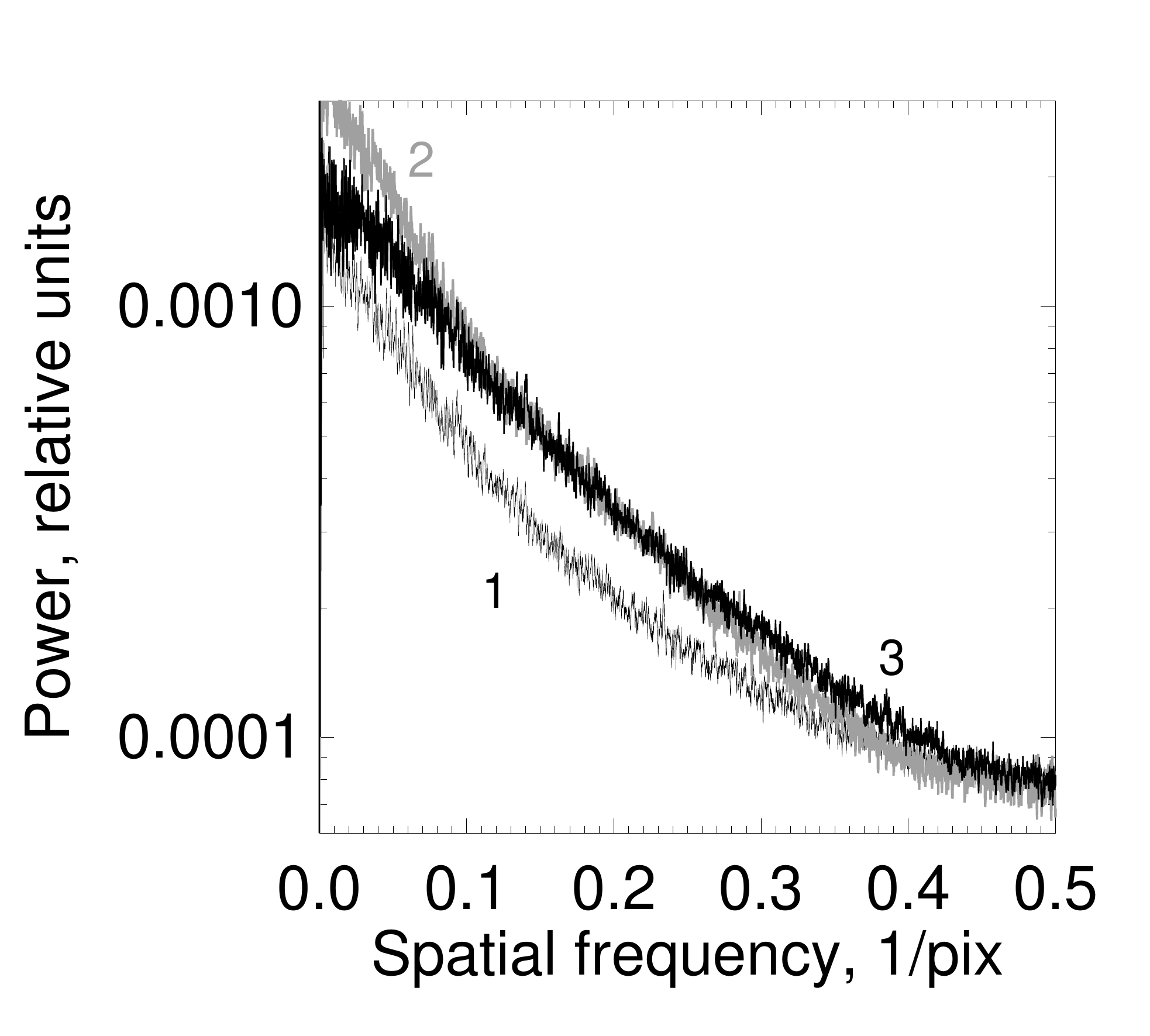}
   \end{tabular}
   \end{center}
   \caption[Figure 8]
   { \label{fig:Figure 8}
 A comparison of power spectra running mean curves for the original image (Figure 7 (a)), stretched image (Figure 7 (b)), and for 27 April 1996 (Figure 1), thin black marked as 1, grey marked as 2, and thick black marked as 3, correspondingly. }
   \end{figure}
 which shows that the 25\,\% stretched image increases the power on the intermediate and low spatial frequencies (compare curves 1 and 2 shown as thin black and grey curves) and leads to about the same power as for the 1996 image (thick black curve marked as 3). A small decrease of the power for the stretched image
(grey) on high spatial frequencies, \textit{e.g.} at $0.4$ pix$^{-1}$ is the result of removing the outer ring with its foreshortened features from the stretched image.

\section{Results }
     \label{S-Results}

Our goal was to compare the spatial ratios for two solar minima, between Cycles 22/23 and Cycles 23/24 and for a number of local minima for 2008\,--\,2011. Figure 9 compares ratios for the Cycles 22/23 \textit{vs.} Cycles 23/24 solar minima and ratios for another pair of quiet-Sun conditions: 1996 \textit{vs.} 2010 with about the same level of SEM EUV flux.
  \begin{figure}[h]
   \begin{center}
   \begin{tabular}{c}
   \includegraphics[height=7.0 cm]{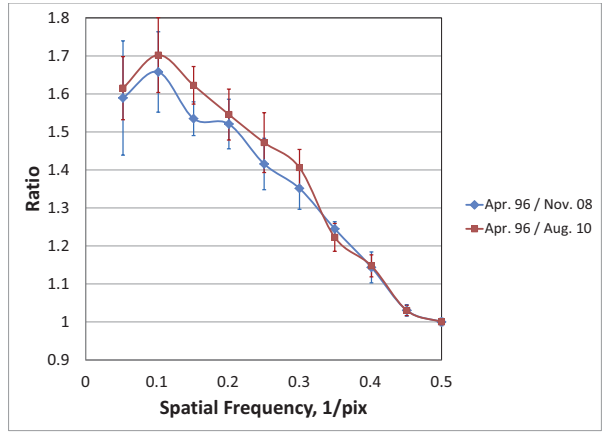}
   \end{tabular}
   \end{center}
   \caption[Figure 9]
   { \label{fig:Figure 9}
 A comparison of power spectra ratios for the two pairs of time intervals, one (diamonds) to compare the global He\,{\sc ii} network structure during two solar minima in April 1996 and in November 2008, another (squares) to compare the April 1996 minimum with the May 2010 local minimum with about the same SEM fluxes (Table 1), 0.94 and 1.01 $10^{10}$ [ph cm$^{-2}$ sec $^{-1}$]. }
   \end{figure}
Figure 9 shows that the smallest change in the network structure between 1996 and either 2008 or 2010 images occurred for small area solar features with spatial scales up to three EIT pixels ($< 5.7$ Mm). For the larger area features ($> 3$ EIT pixels or $> 5.7$ Mm) the ratio curve for 1996 \textit{vs.} 2008 (diamonds) shows lower ratios compared to the 1996 \textit{vs.} 2010 ratios (squares) with the difference outside the error bars for the spatial frequency of 0.15 pix$^{-1}$ (about 6 EIT pixels or about 11.3 Mm). This result shows that the images from the local activity minimum of 2010 contained a smaller number of the features with a spatial size of three to ten pixels (5.7 to 18.8 Mm) than during the Cycles 23/24 minimum and significantly smaller number than during the Cycles 22/23 minimum.

We found small changes in the ratios for the local minima within the current solar cycle, 2008\,--\,2011 (three bottom curves in Figure 4 and three bottom rows in Table 3). They may indicate a number of returns to the conditions of He\,{\sc ii} TR network structure derived from the spatial power spectra for the Cycles 23/24 solar minimum with the increased levels of the SEM 30.4~nm solar irradiance, up to 34\,\% for the minimum of 2011. The ratios for another short time interval (April 1996 \textit{vs.} June 1996) are substantially higher on low and intermediate spatial frequencies than between December 2010 and May 2011 (compare curves marked with diamonds and triangles in Figure 4).

 \section{Concluding Remarks }
     \label{S-Conclusions}

We found that for about the same level of EUV 30.4 nm solar flux from the SEM daily averaged fluxes the ratios for 1996 \textit{vs.} 2010 (see squares in Figure 9) are significantly larger than unity and this appears to be correlated with the change in the distribution of transition-region network structure sizes analyzed for the 2008\,--\,2011 minima. The change of the network structure is consistent with a decreased population of mid-sized features on the solar disk. The same EUV 30.4 nm daily averaged fluxes for these different structures of the transition-region network may be related to the re-distribution of the coronal holes and the larger intensity (contrast) of the small features for the Cycles 23/24 solar minimum and for the local minima of 2010 and 2011 compared to the Cycles 22/23 minimum. These results show that not only the absolute level of EUV irradiance, but its spatial distribution in the transition region, may have to be considered in predicting solar-cycle effects on the thermosphere.

The power spectra and the ratios analyzed in this article show a decrease of the TR spatial He\,{\sc ii} structure
during the prolonged Cycles 23/24 solar minimum. This is consistent with the decrease of the mean TR network
scale \cite{McIntosh11b} determined using the method of ``watershed segmentation'' \cite{McIntosh11a}.
The spectra show \textit{smooth} increases of the power toward the low spatial frequencies
(larger network features). In addition to the results from the method of ``watershed segmentation'' \cite{McIntosh11a}
which provides the histograms with the distribution of the cell radii in the central portion
of the image, the spectra contain a combined effect of the sizes of the structures and their spatial spectral
density. If spatially resolved measurements of He\,{\sc ii} emission are maintained through several solar minima,
we should be able to conclude whether the spatial distribution of that emission provides a proxy for predicting cycles
with anomalously prolonged minima.


\begin{acks}
This work was partially supported by the University of Colorado award 153-5979.
SOHO is a project of international cooperation between NASA and ESA.
\end{acks}


\end{article}

\end{document}